\documentclass[twocolumn]{article}

\usepackage{preprint}

\usepackage{amsmath, amsthm, amssymb, amsfonts}
\usepackage{gensymb}

\usepackage[numbers,square]{natbib}
\usepackage[utf8]{inputenc}	
\usepackage[T1]{fontenc}	
\usepackage{xcolor}		
\usepackage[colorlinks = true,
            linkcolor = purple,
            urlcolor  = blue,
            citecolor = cyan,
            anchorcolor = black]{hyperref}	
\usepackage{booktabs} 		
\usepackage{nicefrac}		
\usepackage{microtype}		
\usepackage{lineno}		
\usepackage{float}			

\usepackage{makecell} %
\usepackage{textcomp}

\usepackage{lipsum}		

\usepackage{newfloat}
\DeclareFloatingEnvironment[name={Supplementary Figure}]{suppfigure}
\usepackage{sidecap}
\sidecaptionvpos{figure}{c}

\usepackage{titlesec}
\titlespacing\section{0pt}{12pt plus 3pt minus 3pt}{1pt plus 1pt minus 1pt}
\titlespacing\subsection{0pt}{10pt plus 3pt minus 3pt}{1pt plus 1pt minus 1pt}
\titlespacing\subsubsection{0pt}{8pt plus 3pt minus 3pt}{1pt plus 1pt minus 1pt}

\usepackage{tikz,xcolor,hyperref}

\definecolor{lime}{HTML}{A6CE39}
\DeclareRobustCommand{\orcidicon}{
	\begin{tikzpicture}
	\draw[lime, fill=lime] (0,0)
	circle [radius=0.16]
	node[white] {{\fontfamily{qag}\selectfont \tiny ID}};
	\draw[white, fill=white] (-0.0625,0.095)
	circle [radius=0.007];
	\end{tikzpicture}
	\hspace{-2mm}
}
\foreach \x in {A, ..., Z}{\expandafter\xdef\csname orcid\x\endcsname{\noexpand\href{https://orcid.org/\csname orcidauthor\x\endcsname}
			{\noexpand\orcidicon}}
}

\title{Ferroelectricity-driven altermagnetism in two-dimensional van der Waals multiferroics}

\usepackage{xwatermark}
\newwatermark[firstpage,color=gray!60,angle=90,scale=0.32, xpos=-4.05in,ypos=0]{\href{https://doi.org/}{\color{gray}{Publication doi}}}
\newwatermark[firstpage,color=gray!60,angle=90,scale=0.32, xpos=3.9in,ypos=0]{\href{https://doi.org/}{\color{gray}{Preprint doi}}}
\newwatermark[
  firstpage,
  color=gray!90,
  angle=0,
  scale=0.28,
  xpos=0in,
  ypos=-5in
]{%
*\,\texttt{renwei@shu.edu.cn}\\[4pt]%
\textdagger\,\texttt{haowang@tmm.tu-darmstadt.de}%
}

\usepackage{authblk}
\makeatletter
\renewcommand\AB@affilsepx{\protect\\} 
\makeatother
\author[1]{Bo Zhao}
\author[1]{Fu Li}
\author[2\thanks{\tt{renwei@shu.edu.cn}}]{Wei Ren}
\author[1\thanks{\tt{haowang@tmm.tu-darmstadt.de}}]{Hao Wang}
\author[1]{Hongbin Zhang}

\affil[1]{Theory of Magnetic Materials, Institute of Materials Science, Technical University of
Darmstadt, 64287 Darmstadt, Germany}
\affil[2]{Institute for Quantum Science and Technology, State Key Laboratory of Advanced Refractories, Materials Genome Institute, Physics Department, Shanghai University, Shanghai 200444, China}

\begin{document}

\twocolumn[ 
  \begin{@twocolumnfalse} 

\maketitle

\begin{abstract}
Altermagnets (AMs) are a recently identified class of unconventional collinear compensated antiferromagnets that exhibit momentum-dependent spin splitting despite having zero net magnetization. This unconventional magnetic order gives rise to a range of phenomena, including the anomalous Hall effect, chiral magnons, and nonlinear photocurrents. Here, using spin space group (SSG) symmetry analysis and first-principles calculations, we demonstrate an efficient strategy to control altermagnetism in two-dimensional multiferroics through ferroelectric polarization and interlayer sliding. For material realization, we find that monolayer and bilayer FeCuP$_2$S$_6$ exhibit finite spin splitting when ferroelectric sublattices are connected by nonsymmorphic screw-axis operations rather than pure translation or inversion symmetry. Interlayer sliding further enables reversible switching or suppression of spin splitting through modifications of the SSG. Our calculations further reveal that the anomalous Hall response serves as a direct probe of these spin-split states. These findings establish two-dimensional van der Waals multiferroics as promising platforms for realizing electrically controllable altermagnetism and advancing next-generation spintronic and magnetoelectric technologies.
\end{abstract}
\vspace{0.35cm}

  \end{@twocolumnfalse} 
] 



\section{Introduction}
Spintronics exploits the electron's spin degree of freedom to process and store information~\cite{sinova2015spin,hirsch1999spin,li2025high}. The underlying magnetic order, described in Landau’s framework by a local order parameter, is traditionally categorized into ferromagnetism (FM) and antiferromagnetism (AFM). FM provides net spin polarization and external controllability, whereas AFM, defined by the Néel vector, exhibits ultrafast dynamics and robustness against stray fields~\cite{baltz2018antiferromagnetic}. Recently, a new unconventional magnetic paradigm, altermagnetism (AM)~\cite{krempasky2024altermagnetic,song2025altermagnets,chen2025unconventional}, has been proposed. It is analogous to even-parity wave superconductivity~\cite{schofield2009there,vsmejkal2022emerging}, exhibiting collinear-compensated order with momentum-dependent band splitting. The conventional magnetic space group (MSG) framework cannot capture this behavior because of the unlocked spin and lattice symmetries in the weak spin-orbit coupling (SOC) limit. To properly describe such systems, one must instead use the spin space group (SSG) formalism.
This new class of magnets effectively combines key features of FM and AFM orders, giving rise to a rich variety of emergent phenomena, including anomalous Hall effects~\cite{zhou2024crystal,feng2022anomalous,gonzalez2023spontaneous,reichlova2024observation}, chiral magnons~\cite{hoyer2025spontaneous}, magneto-optical Kerr responses~\cite{song2025altermagnets,iguchi2025magneto}, and nonlinear photocurrents~\cite{jiang2025nonlinear}.

Harnessing the unique advantages of altermagnetism has become a central challenge for next-generation spintronic devices. In general, realizing ``electric write and magnetic read" functionality enables device miniaturization and higher integration density~\cite{zutic2019spintronics}.
The most straightforward route is to manipulate magnetism by applying an external electric field~\cite{chang2016discovery,gao2019new}; however, this approach often suffers from significant energy dissipation and unfavorable device scaling~\cite{jungwirth2016antiferromagnetic,han2024electrical}. In contrast, magnetoelectric coupling achieved through ferroelectric switching offers a much more efficient alternative~\cite{gu2025ferroelectric,liu2020magnetoelectric, zhu2025sliding,sun2025proposing,wang2025two}. Strong magnetoelectric coupling is typically found in multiferroic materials. However, such systems are rare in nature because ferromagnetism, which usually occurs in metals with partially filled $d$ orbitals, is intrinsically incompatible with ferroelectricity, which requires insulating states. These challenges underscore the need for alternative routes to magnetoelectric coupling. 

Systems with inherently separable spin and lattice symmetries—such as altermagnets—offer a natural solution~\cite{duan2025antiferroelectric,wang2024electric,wang2025two, camerano2025multiferroic, zhang2024predictable}. In addition, a growing number of studies~\cite{vizner2021interfacial,yang2018origin,he2023nonrelativistic,liu2024twisted,wei2024plane,wei2025sliding,mavani2025two} have shown that sliding ferroelectricity can also serve as an effective mechanism to tune magnetic order. Yet, the mechanism in non-relativistic AMs remains largely unexplored, since their disentangled spin-lattice symmetry demands a SSG description~\cite{jiang2024enumeration,vsmejkal2022beyond,chen2024enumeration,xiao2024spin} beyond the conventional MSG. The role of interlayer coupling and its impact on magnetic symmetry therefore remain open questions, and suitable material platforms for realizing this coupling are still lacking.

In this letter, we propose a general strategy to control magnetic order in altermagnetic systems via ferroelectric polarization. Our study focuses on the interplay between SSG symmetry and band splitting, and demonstrates how interlayer sliding can effectively tune the spin splitting. Specifically, in the two-dimensional van der Waals multiferroic FeCuP$_2$S$_6$, the antiferroelectric antiferromagnetic (AFE–AFM) order is not connected by a simple inversion or translation but by a nonsymmorphic screw-axis operation, which protects the emergence of the AM state. In the AA-stacked bilayer, when the structure possesses a screw axis along the b direction, AM appears at both $\vec{b} = 0$ and $\vec{b} = 1/2$, where it is preserved by screw and rotation symmetry, respectively. Sliding along the screw axis thus toggles the spin splitting on and off, which can be confirmed by the calculated shift current and anomalous Hall effect (AHE). These results establish FeCuP$_2$S$_6$-type AFE–AFM systems as ideal platforms for exploring magnetoelectric-driven altermagnetism.


\section{Results and discussion}

First, we understand AM from the perspective of the SSG. Recently, several works have systematically enumerated the SSGs~\cite{vsmejkal2022emerging,jiang2024enumeration,chen2024enumeration,xiao2024spin}. The basic idea is to separate the spin degree of freedom from the lattice symmetry and use group extension to form a larger group that includes their joint operations. In the conventional space group (SG) $G$, the translation group $T$ is an invariant subgroup, and the quotient group $G/T$ is isomorphic to a point group $P$, i.e., $G/T \cong P$. Owing to the constraint imposed by the lattice periodicity, the allowed point groups must belong to the 32 crystallographic point groups. For the MSG, since the spin moment can be regarded as an axial vector, the symmetry operator should be written as $U = \det (R)R$. Additionally, the effect of time-reversal symmetry is simply to flip the spin orientation, which can be denoted by $1'$, forming a two-element group $R=\{1, 1'\}$, commonly referred to as the time-reversal or spin-reversal group. Thus, the MSG can be viewed as the outer direct product of an SG and the time-reversal group $R$, representing by a double group. Depending on how the spatial operations combine with time reversal, MSGs can be classified into the four Shubnikov types.

In essence, an MSG is a space group extended by a binary group, meaning that the lattice and spin degrees of freedom are locked together — a situation valid in the presence of strong SOC. However, to describe systems where the spin and lattice are only weakly coupled, a more general framework is required. 

The operators in SSGs are represented as \{$\mathcal{O} || \mathcal{R} $\}, with the operator $\mathcal{O}$ applying to the spin and $\mathcal{R}$ applying to the lattice individually.
\citeauthor{jiang2024enumeration}~\cite{jiang2024enumeration} enumerated SSGs ($G^{(s)}$) based on the invariant subgroups of SGs, with the spin part treated as a three- dimensional (3D) real representation of the quotient groups $Q$, which is given by 
\begin{equation}
    Q = G/H \cong S/S_0,
\end{equation}
where $G$ represents the lattice part, $H$ corresponds to the pure lattice subgroup, $S$ denotes the spin part, and $S_0$ forms the pure spin subgroup. This means that, after excluding pure lattice/spin operations,
the remaining parts of the lattice/spin part are both isomorphic to a point group. Unlike in space groups, where the invariant subgroup is the translation group $T$, the invariant subgroup $H$ in SSGs can include combined rotation-translation operations. Consequently, the quotient group can be either crystallographic or noncrystallographic. According to the symmetry of the pure spin subgroup $S_0$, magnetic systems can be classified into collinear, coplanar, and general noncoplanar cases. 

For the collinear case $S_0 = \{C_{\theta} || E|\mathbf{0}\} + \{M_xC_{\theta} || E|\mathbf{0}\}$, the quotient group $Q=G^{(s)}/S_0$ can be isomorphic to $C_1 = \{E \}$ or $C_s = \{E, \mathcal{T}\}$. When $Q\cong C_1$, the spin configuration is unique, corresponding to the conventional ferromagnetic order. In the case of $Q \cong C_s$, we need to consider whether $Q\cong T/T_H$, the translation quotient group, or $P/P_H$, the point quotient group.
If $Q\cong T/T_H$, the system must extend its primitive cell into a supercell to restore the spin configuration, corresponding to the traditional antiferromagnetic order. If $Q\cong P/P_H \cong C_s$, the SSG can be expressed as follows: $G^{(s)} = H + \{ M_z || \mathcal{RT}|\mathcal{\tau}\}H$. When $\mathcal{R} = \mathcal{P}$, there exists a combined spatial inversion and time-reversal symmetry $\mathcal{PT}$, which enforces spin degeneracy in the whole Brillouin zone, describing a conventional antiferromagnet. In contrast, if $\mathcal{R}\neq \mathcal{P}$, i.e., the operation is not a spatial inversion; no symmetry protects the global spin degeneracy — this corresponds to altermagnetism.

In short, from the viewpoint of spin-group symmetry, the emergence of AM requires a collinear magnetic configuration, where the quotient group $Q$, formed from the spin part over the pure spin group,
is isomorphic to the binary group $\mathbb{Z}_2^{\mathcal{T}}$. Importantly, the point-group quotient group $P/P_H$, which is also isomorphic to $Q$, must not originate solely from lattice translations, nor involve the combined $\mathcal{PT}$ symmetry.

\begin{figure}[H]
  \centering
  \includegraphics[width=0.45\textwidth]{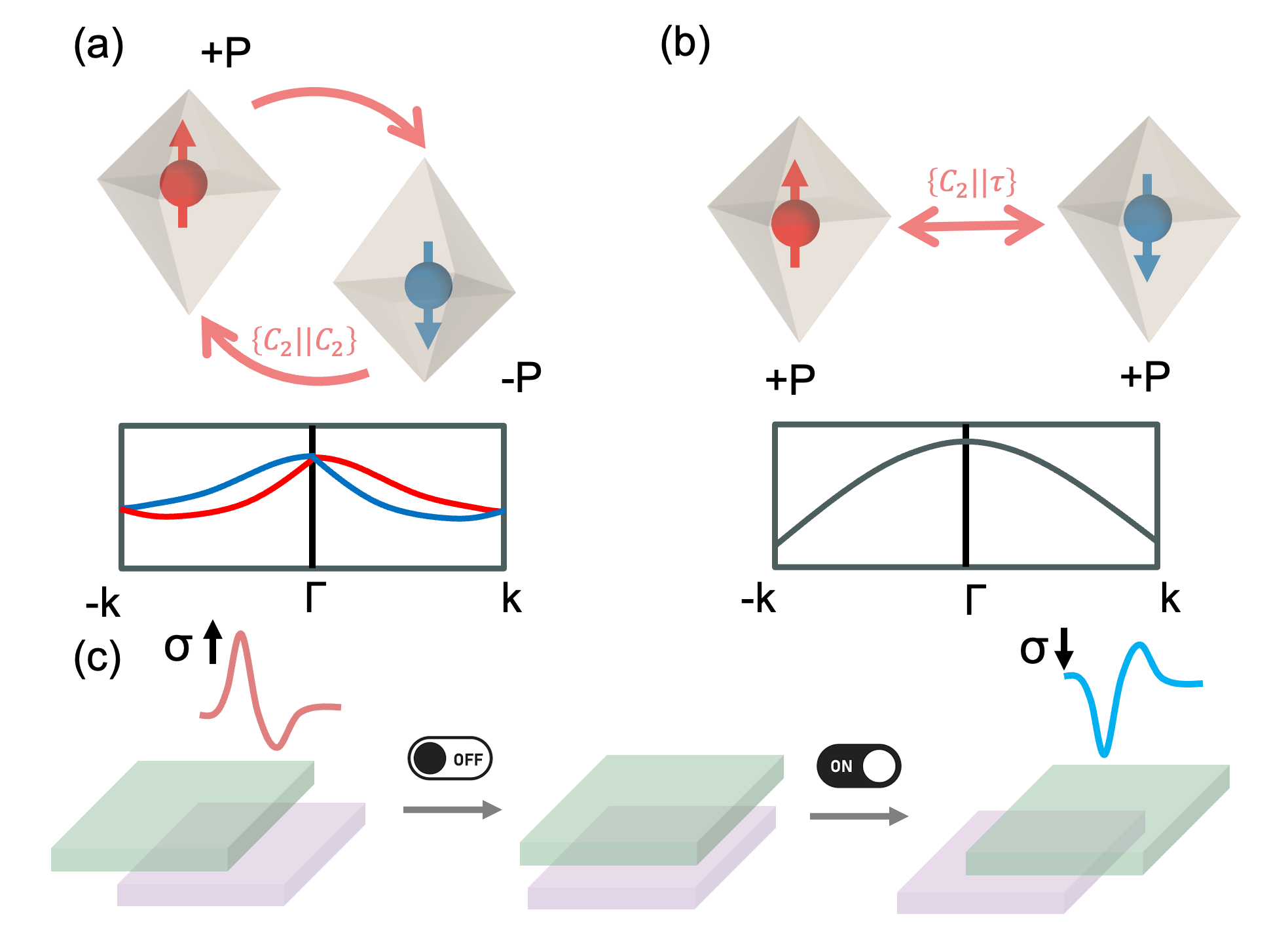}
  \caption{Schematics of two FE sublattices with opposite spins in (a) AFEAFM and (b) FEAFM structures. Arrows indicate the transformation between them, where in AFEAFM the lattice part is connected by $C_2$ rotation and in FEAFM it is connected by translation. Spin splittings can be observed in AFEAFM band structure, while degenerated in the FEAFM bands. (c) Schematics of switching spin splitting through bilayer sliding. $\sigma$ stands for the inverse Berry-curvature related properties.
  }
  \label{fig:fig1}  
\end{figure}

Following this criterion,  \citeauthor{duan2025antiferroelectric}~\cite{duan2025antiferroelectric} proposed that antiferroelectric (AFE) systems provide an ideal playground for realizing AM. In the ferroelectrically coupled antiferromagnetic (FEAFM) phase, as shown in Figure~\ref{fig:fig1}(b), each ferroelectric lattice acts as a structural block, and adjacent sublattices are connected by pure translations, leading to $T/T_H\cong C_s$ and a conventional AFM configuration. In contrast, in the antiferroelectrically coupled antiferromagnetic (AFEAFM) phase (Figure~\ref{fig:fig1}(a)), the translational symmetry is broken, and the sublattices are related through a symmorphic or nonsymmorphic operation $\mathcal{R}$. Such reduced symmetry enables spin splitting along specific {k}-paths in momentum space. Moreover, because the Berry curvature is a pseudovector that transforms in the same way as the spin, the on/off switching of spin splitting among distinct bilayer stackings leads to an incomplete cancellation of Berry curvature, thereby producing a finite anomalous Hall conductivity $\sigma_{AH}$, as illustrated in Figure~\ref{fig:fig1}(c).

\begin{figure*}[t]
  \centering
  \includegraphics[width=0.67\textwidth]{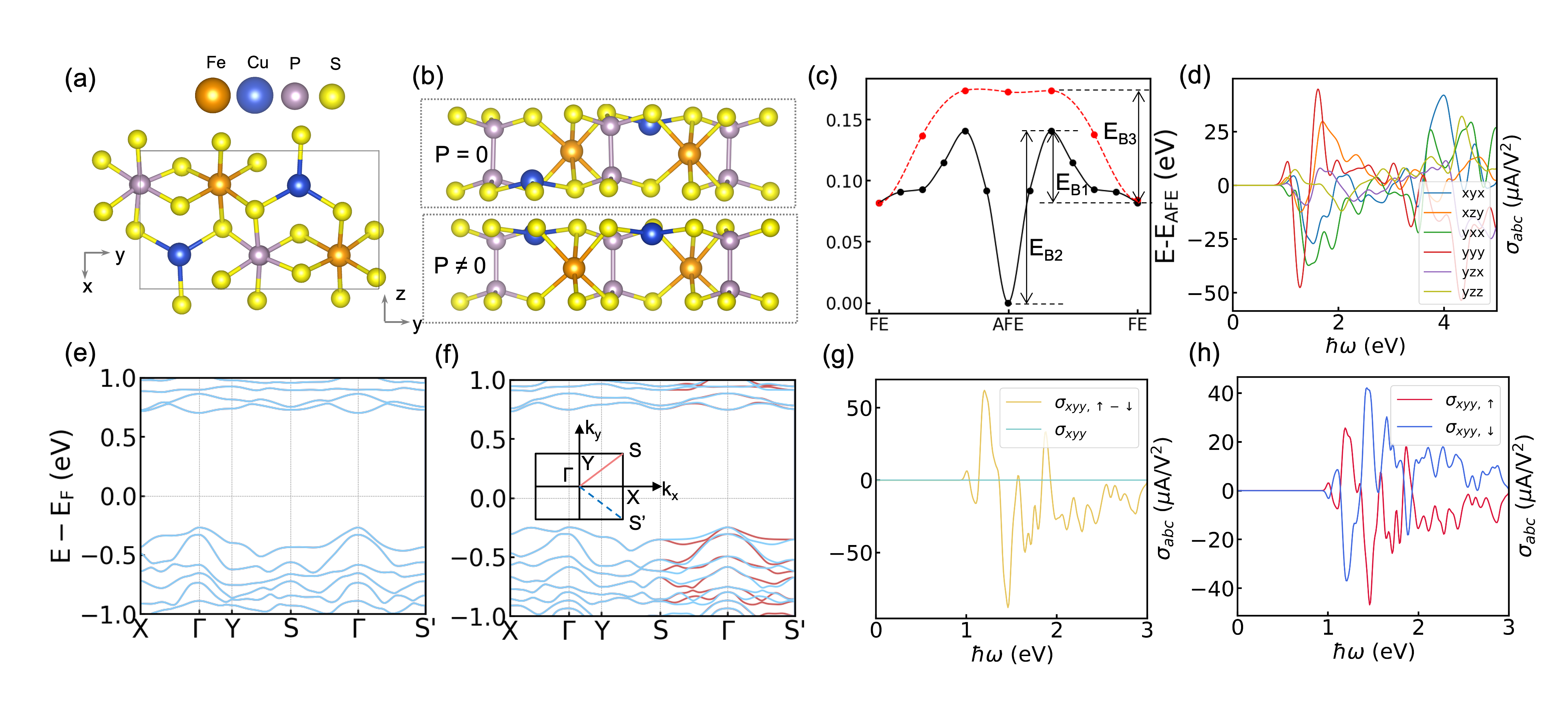}
  \caption{(a) Top view of antiferroelectric monolayer (AFE-ML) FeCuP$_2$S$_6$ structure, the orange, blue, light purple, and yellow balls stand for Fe, Cu, P, and S atoms, respectively. (b) Side view of the AFE-ML FeCuP$_2$S$_6$ (up) and the corresponding FE-ML configuration (down), respectively. (c) The transition pathways between two FE states (positive and negative polarization P) through PE state (red dashed line) and through AFE state (black line). The energy barriers of the FE-to-AFE, AFE-to-FE and FE-to-FE are denoted as E$_{B1}$, E$_{B2}$, and E$_{B3}$, respectively. (e) and (f) are the band structures of FE- and AFE-ML FeCuP$_2$S$_6$, respectively. The Brillouin zone and the high-symmetry $k$-path are shown in the inset of (d). (d) Shift current tensor for AFE-ML FeCuP$_2$S$_6$. (g) The $\sigma_{xyy}$ component of the charge shift current and the spin current tensor for AFE-ML FeCuP$_2$S$_6$, respectively. (h) The spin up and down channels of the spin shift current component $\sigma_{xyy}$.
  }
  \label{fig:fig2}  
\end{figure*}

Based on the above analysis, for the material realization, the ideal targeted ferroelectric-driven AM materials should satisfy the three essential criteria: (1) an AFE ground state or a ferroelectric (FE) phase in which lattice distortions generate the requisite $\mathcal{R}$ symmetry connecting the magnetic sublattices~\cite{zhu2025two,zhu2025emergent}, (2) with weak SOC, (3) a moderate ferroelectric transition barrier. The FePX$_3$(X = S, Se, and Te) families~\cite{li2013coupling,chittari2016electronic,ding2017prediction} fulfills these requirements and stands out as a promising candidate~\cite{sun2025proposing,wang2025magnetic}. These compounds belong to a two-dimensional (2D) van der Waals type-III multiferroics, where ferroelectric and ferromagnetic order parameters are intimately coupled by symmetry. The crystal structure of FePS$_3$ is shown in Figure~\ref{fig:fig2}(a), which has been predicted to possess a FEAFM ground state~\cite{wang2025magnetic}. Substituting other metal atoms on the Fe site provides an effective means to tune both the ferroelectric and magnetic orders.
In the MPX$_3$ family, heterovalent substitution forming $\mathrm{M_I}\mathrm{M_{III}}$P$_2$S$_6$ ($\mathrm{M_I}$ = Ag$^+$, Cu$^+$; $\mathrm{M_{III}}$ = In$^{3+}$, V$^{3+}$, Fe$^{3+}$, etc.) drives opposite vertical displacements of cations, resulting in distinct orientations of ferroelectric polarization~\cite{wang2018new}.

Here we focus on the monolayer (ML) FeCuP$_2$S$_6$, whose FE and AFE structures shown in Figure~\ref{fig:fig2}(b). The FE-ML structure belongs to $P1$ space group, while the AFE-ML adopts $P2_1$, which contains a two-fold screw operation along y axis. Following the notation of \citeauthor{jiang2024enumeration}~\cite{jiang2024enumeration}, the spin space group of FE and AFE can be identified as 1.2.1.1.L and 4.1.2.1.L, respectively. For 4.1.2.1.L, the number of elements in the point and translational quotient groups are $I_k = P/P_H = 1$ and $I_t = T/T_H = 2$, respectively. Consequently, this SSG consist of two elements, $\{E || E|0 \}$ and $\{m_z || C_2|0\frac{1}{2}0 \}$. The spin-unlocked operation $\{m_z || C_2|0\frac{1}{2}0 \}$ decouples the spin and the lattice parts, and resulting in a spin-splitting bands along $\Gamma$ - S path in the BZ, as shown in Figure~\ref{fig:fig2}(f). In contrast, the FEAFM phase possesses SSG operations $\{E || E|0 \}$ and $\{m_z || E|001 \}$, corresponding to the type-III MSG 1.3.3 and exhibiting conventional AFM behavior shown in Figure~\ref{fig:fig2}(e). 

Since the ferroelectric polarization can be switched between the FE and AFE states by an external electric field, we evaluated the stability and the transition barrier between these phases. Figures~\ref{fig:fig2}(a) and (b) show the FE, AFE, and paraelectric (PE) configurations, and the corresponding transition pathway is illustrated in Figure~\ref{fig:fig2}(c). The energy profile indicates that the ground state is AFEAFM, which lies 81.5 meV/f.u. lower in energy than the FEAFM phase. An FE-AFE transition can occur when the external bias $E_{FE-AFE}$ satisfies $E_{B1}<E_{FE-AFE}<E_{B3}$, while the reverse FE-AFE transition happens when the external field $E_{AFE-FE}$ > $E_{B2}$. $E_{B1}$, $E_{B2}$, and $E_{B3}$ correspond to the barriers for FE $\rightarrow$ AFE, AFE $\rightarrow$ FE, and FE $\rightarrow$ PE $\rightarrow$ FE transitions, respectively. For ML FeCuP$_2$S$_6$, the calculated $E_{B1}$, $E_{B2}$, and $E_{B3}$ are 60, 140 and 90 meV/f.u., respectively. It reveals the capability for experimental realization~\cite{sui2023sliding,ding2017prediction}. Notably, owing to the screw-axis symmetry, the AFE-ML phase is also non-centrosymmetric and exhibits a band gap of about 1 eV, enabling a steady photocurrent response. The symmetry-restricted nonzero components of the charge shift current tensor for AFE-ML FeCuP$_2$S$_6$ are shown in Table~S2. Figure~\ref{fig:fig2}(d) presents the six dominant components of the shift-current tensor, among which the dominant term $\sigma_{yyy}$ yields a shift current of –48.5 $\mu$A/V$^2$ at $\hbar\omega$ = 1.26 eV, indicating its potential applicability in the bulk photovoltaic effect (BPVE)~\cite{fridkin2001bulk,yang2024two,wei2025shift}. 
Beyond the conventional charge shift current, the AFE-ML phase also allows for a spin-resolved shift response~\cite{gu2025ferroelectric,young2013prediction,xu2021pure,dong2025crystal}. For the charge shift current, certain tensor components such as $\sigma_{xxx}$ and $\sigma_{xyy}$, are strictly forbidden by the crystal-symmetry constraints of the third-rank shift-current tensor and therefore vanish. However, in the altermagnetic state governed by SSG symmetry, the band structure becomes spin-split along the $\Gamma - S$ and $\Gamma - S'$ paths. As a result, these ``forbidden” components can accumulate a pure spin flow rather than a net charge flow, giving rise to the nonzero spin shift current, defined as~\cite{young2013prediction,xu2021pure}
\begin{equation}
    \sigma_{ijk}^{\mathrm{spin}} = \sigma_{ijk}^{\uparrow} - \sigma_{ijk}^{\downarrow}
\end{equation}

We explicitly compute the spin-resolved shift current for AFE-ML FeCuP$_2$S$_6$, as shown in Figure~\ref{fig:fig2}(g). While the charge response $\sigma_{xyy}$ is strictly zero, the opposite contributions of $\sigma_{xyy}^{\uparrow}$ and $\sigma_{xyy}^{\downarrow}$ yield a finite $\sigma_{xyy}^{\mathrm{spin}}$. A dc spin current emerges once the photon energy exceeds the band gap and reaches its first peak of 6.5 $\mu$A/$V^2$. This demonstrates that in altermagnetic FeCuP$_2$S$_6$, a pure spin current generated by linearly polarized light arises from the underlying quantum geometry and symmetry of the spin-split bands, rather than from a conventional SOC-driven spin Hall conductivity. This highlights that two-dimensional AMs provide a symmetry-protected platform for experimentally detecting pure spin currents through nonlinear optical measurements.

\begin{figure*}[t]
  \centering
  \includegraphics[width=0.67\textwidth]{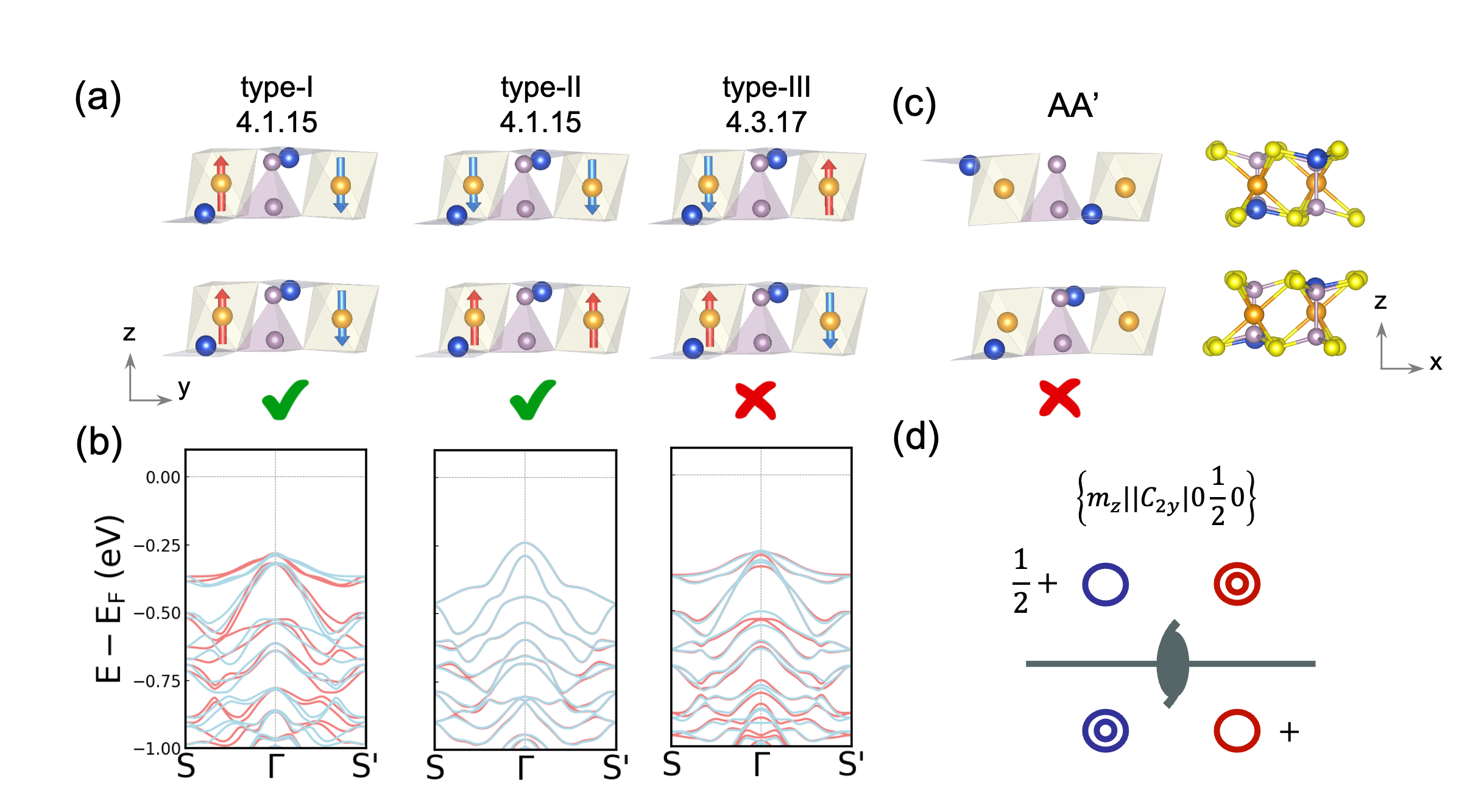}
  \caption{(a) Three possible stacking configurations for BL FeCuP$_2$S$_6$and the corresponding MSG. Type-I and -II exhibit altermagnetism, while type-III shows trivial compensated ferromagnetism; The corresponding band structures along the spin-splitting $k$-path are shown in (b); (c) Side views of AA' stacking mode of FeCuP$_2$S$_6$.
  (d) Schematics for the minimal symmetry requirement for AM. Open circles and double-ring circles are symmetry equivalent sites, respectively. Red and blue colors represent different spin up and spin down, respectively.
  }
  \label{fig:fig3}  
\end{figure*}

Since monolayer FeCuP$_2$S$_6$ is a van der Waals (vdW) material, in addition to applying an external electric field, tuning the interlayer ferroelectric polarization through layer sliding provides an efficient way to control its ferroic order. To explore whether interlayer sliding can serve as an effective knob to modulate the AM state, we constructed a bilayer (BL) FeCuP$_2$S$_6$ with AA stacking, where the top layer is directly aligned with the bottom one. As shown in Figure~\ref{fig:fig3}(a), three stacking types are considered: type-I and type-III are AFM within each layer, while type-II is FM intralayer. For type-I and type II, the SG and the SSG are identified as P2$_1$ (4.1.2.L), in which the two magnetic sublattices are related by the operation $\{m_z||C_{2y}|0\frac{1}{2}0\}$. This symmetry is equivalent to that of the monolayer AFEAFM phase. Consequently, the existence of the screw-axis symmetry is expected to induce altermagnetism, which is indeed confirmed by the spin splitting along the $\Gamma$-S path shown in Figure~\ref{fig:fig3}(b) (the full Brillouin zone band structures are provided in Figure-S5). 

Notably, in the type-I stacking, each monolayer itself already hosts an AM state, and the stacking preserves the screw-axis symmetry; thus, the overall bilayer remains altermagnetic. In contrast, the type-II stacking consists of AFEFM monolayers that are non-altermagnetic individually, but the interlayer stacking restores the screw-axis symmetry, and thus an AM state emerges. In this configuration, the AFM sublattices of Fe are coupled only through interlayer interactions, leading to a weaker spin splitting compared with type-I. Interestingly, a finite magnetic moment of approximately 0.085 $\mu_B$ is observed on the Cu atoms. In this case, the magnetic contribution of Cu must be considered in understanding the AM behavior of the type-II stacking. According to the symmetry analysis, the Wyckoff positions of Cu still belong to a point group containing the $2_1$ operation, identical to that of Fe. Therefore, despite the nonzero magnetic moment on Cu, the overall SSG symmetry remains intact, and the altermagnetic state is preserved. Furthermore, the spin- and orbital-resolved band structures (Figure~S6 and s7) reveal that in both type-I and type-II stackings, the valence bands are mainly derived from Cu \textit{d} orbitals, while both Cu and Fe contribute to the spin-split bands. This indicates that the emergence of altermagnetism is primarily governed by symmetry: even when local magnetic moments are small or sublattices are coupled only weakly through interlayer interactions, spin splitting can still be clearly observed in the electronic band structure.

To further understand the origin of the Cu magnetic moments in the type-II stacking, we analyze the interaction based on the Goodenough-Kanamori rules~\cite{kanamori1959superexchange,sivadas2018stacking,zhang2019intrinsic}. The ferromagnetic (FM) coupling between Fe and Cu arises from the Fe-ligand-Cu superexchange interaction, where the Fe-S-Cu bond angles are close to 90$\degree$. Due to this orthogonality, the partially occupied $d$ orbitals of Fe and Cu hybridize with different $p$ orbitals of the bridging S atoms. According to Hund’s rule, parallel spins can hop through these two bonds, rendering the Cu atoms weakly ferromagnetic. In this configuration, Cu is simultaneously coupled to both the magnetic and ferroelectric orders, implying a modified magnetoelectric coupling compared with systems consisting of intralayer AFM Fe atoms and nonmagnetic Cu atoms.

The type-III stacking, on the other hand, corresponds to the magnetic space group (MSG) 4.3.17 and the spin space group (SSG) 4.1.1.1.L, belonging to a type-III MSG, where the spin and lattice degrees of freedom are synchronized through identical rotational operations. The calculated band structure (Figure~\ref{fig:fig3}(b)) shows spin degeneracy breaking along the $\Gamma$-S path, while no symmetry-protected spin splitting appears along symmetry-equivalent or other $k$-paths (see Figure~S5), indicating a compensated magnetic state. 
This spin splitting is $k$-independent and originates solely from the exchange (Zeeman-type) interaction. A similar splitting at the $\Gamma$ point was reported in Ref.~\cite{yuan2024nonrelativistic} and attributed to the SOC-induced change of the MSG, caused by the absence of a symmetry relating the two sublattices. Figure~\ref{fig:fig3}(d) summarizes the minimal symmetry requirement: in a bilayer system, spin splitting emerges when two magnetic atoms located along the main screw axis possess opposite spin orientations. For the type-II stacking, this condition also applies to the magnetic sublattices of Cu atoms, which carry finite magnetic moments.

Another low-energy stacking mode is AA$'$ stacking, which is constructed by a 180\degree rotation about z axis of the A layer, as illustrated in Figure~\ref{fig:fig3}(c). The SG in this case reduces to $P1$, and correspondingly, the SSG of the three types of stackings proposed above includes symmetry of either $\{ C_2 ||E \}$ or $\{ C_{\theta\infty} ||E \}$. No symmetry operation is present to protect the spin degeneracy; therefore, the AA$'$-stacked BL FeCuP$_2$S$_6$ turns out to be compensated ferromagnetism with spin-split bands observed in the full BZ as shown in Figure~S5.

\begin{figure*}[!t]
  \centering
  \includegraphics[width=0.67\textwidth]{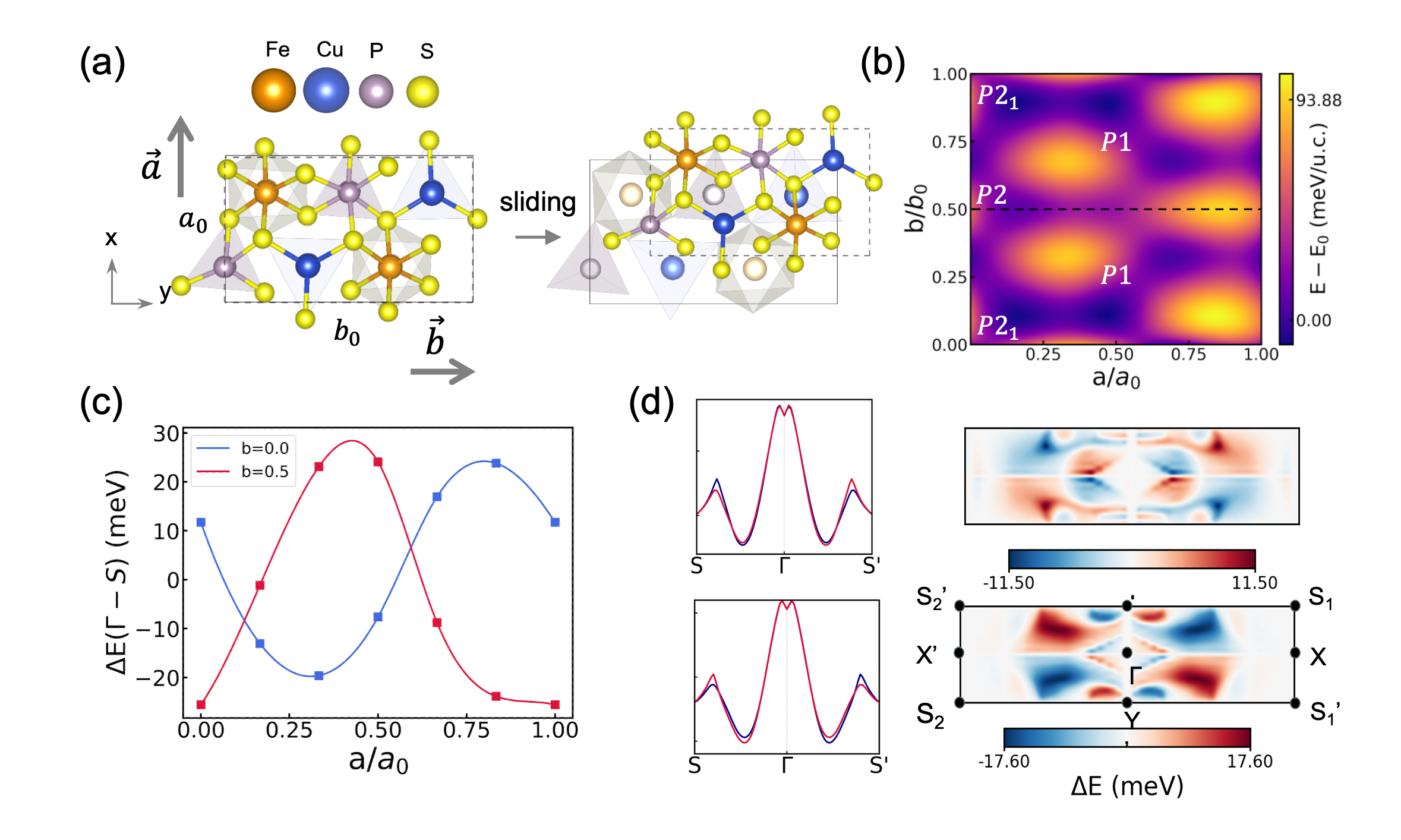}
  \caption{(a) The top view of the AA stacking BL FeCuP$_2$S$_6$ (left) and a sliding configuration along x-y plane (right). (b) The sliding potential energy surface of AA BL stacking, number of the space group is marked at the high-symmetry points. (c) spin-splitting energy $\Delta E$ upon sliding along $\vec{a}$ direction, with $\vec{b}$ at $\vec{b}/b_0 = 1$ (blue) and $\vec{b}/b_0 = \frac{1}{2}$ (red); (d) Reversed spin splittings observed in the bands at 0.75 eV below $E_F$ in Type II BL at (top) $\vec{a}/a_0 = \vec{b}/b_0 = 1$ and (bottom) $\vec{a}/a_0 = \vec{b}/b_0 = \frac{1}{2}$. The corresponding spin-splitting energy heat maps on $k_z = 0$ plane are shown on the right.
  }
  \label{fig:fig3b}  
\end{figure*}


Following the high-symmetry cases, we investigate how interlayer sliding at generic positions influences altermagnetic behavior. As shown in Figure~\ref{fig:fig3b}(a), we apply sliding along the $\vec{a}$ and $\vec{b}$ directions, for the type-I and -II of bilayer FeCuP$_2$S$_6$. We first calculate the sliding energy surface shown in Figure~\ref{fig:fig3b} (b). The calculated sliding energy shows a variation within 100 meV/u.c., demonstrating that ferroelectric sliding is energetically feasible for experimental realization. To quantitatively describe the relationship between sliding and the spin splitting, we define the magnitude of the spin splitting along a specific $k$-path $\Delta$ as 
\begin{equation}
    \Delta E (\Delta) = \frac{1}{N}\frac{1}{N_k}\sum_{n=1}^N\sum_{k, k\in \Delta}^{N_k}[\epsilon_{n,\downarrow}(k) - \epsilon_{n,\uparrow}(k)],
\end{equation}
where $\epsilon_{n,\uparrow}(k)$ and $\epsilon_{n,\downarrow}(k)$ are the spin-up the the spin-down energy of band $n$. As seen in Figure~\ref{fig:fig3b}(b), multiple local minima appear on the sliding-energy surface. For example, along the $\vec{b}$ direction at $\vec{a}/a_0 = 0.125$, energy minima occur at $\vec{b}/b_0$ = 0.125, 0.5, and 0.75. The results also indicate that sliding along $b$ axis changes the symmetry of the bilayer, whereas sliding along $a$ at fixed $b$ preserves it. At $\vec{b}/b_0=1$, the structure reaches a high-symmetry point with SG and SSG being $P2_1$ and 4.1.2.1.L, identical to those of the three initial stacking types discussed earlier. When $\vec{b}/b_0=1/2$, the SG and SSG become $P2$ and 3.1.2.L; here, the lattice symmetry contains a  two-fold rotation $C_{2y}$ instead of the screw operation which does not exist in the original ML and AA-stacked BL structures, and the operation $\{ m_z ||C_2 \}$ preserves the \textit{k}-dependent spin splitting. At the other sliding positions, the two-fold symmetry is broken and the space group reduces to $P1$. The emergence of altermagnetism thus strictly follows the symmetry condition observed in the high-symmetry cases -- AM occurs when $\{ C_{2y} | 0\frac{1}{2}0 \}$ or $C_{2y}$ is present, i.e., $\vec{b}/b_0=0.0$ and $\vec{b}/b_0=1/2$, independent of the $a$ displacement.

Figure~\ref{fig:fig3b}(c) shows the spin-splitting energy of type-I stacking for sliding along the $\vec{a} + b_0$ and $\vec{a} + \frac{1}{2}b_0$ directions. In both sliding paths, the sign of the spin-splitting energy reverses the sign with a varying value of $\vec{a}$. Additionally, as shown in Figure~\ref{fig:fig3b} (c), reversed spin polarization is also observed for $\vec{b}/b_0=1$ and $\vec{b}/b_0=1/2$ at given $\vec{a}$. It demonstrates that the switch of the spin polarization of the path can be realized either by a change of the structural symmetry or by variation of the interlayer coupling, both without changing the N\'eel vector. The two spin splitting energy curves $\mathrm{\Delta E(\Gamma - S)}$ (Figure~\ref{fig:fig3b}(c)) are not strictly symmetric with respect to vector $\vec{a}$. This is because there exists only uniaxial symmetry along y-axis, while the S atoms are distorted along x-axis. To further investigate the distribution of spin splitting over the BZ, we calculate $\Delta E$ on $k_z = 0$ plane of the BZ and plot as contour figures for two spin-splitting reversed bands driven by sliding from $\vec{a}/a_0 = \vec{b}/b_0 = 1$ to $\vec{a}/a_0 = \vec{b}/b_0 = \frac{1}{2}$ (Figure~\ref{fig:fig3b}(d) top and bottom panel, respectively). The corresponding heatmap in Figure~\ref{fig:fig3b}(d) shows that spin splitting occurs across the $k_z=0$ plane except the high symmetry paths $\Gamma$-$X$ and $\Gamma$-$Y$ which are protected by $\mathcal{PT}$ symmetry.

\begin{figure}[H]
  \centering
  \includegraphics[width=0.49\textwidth]{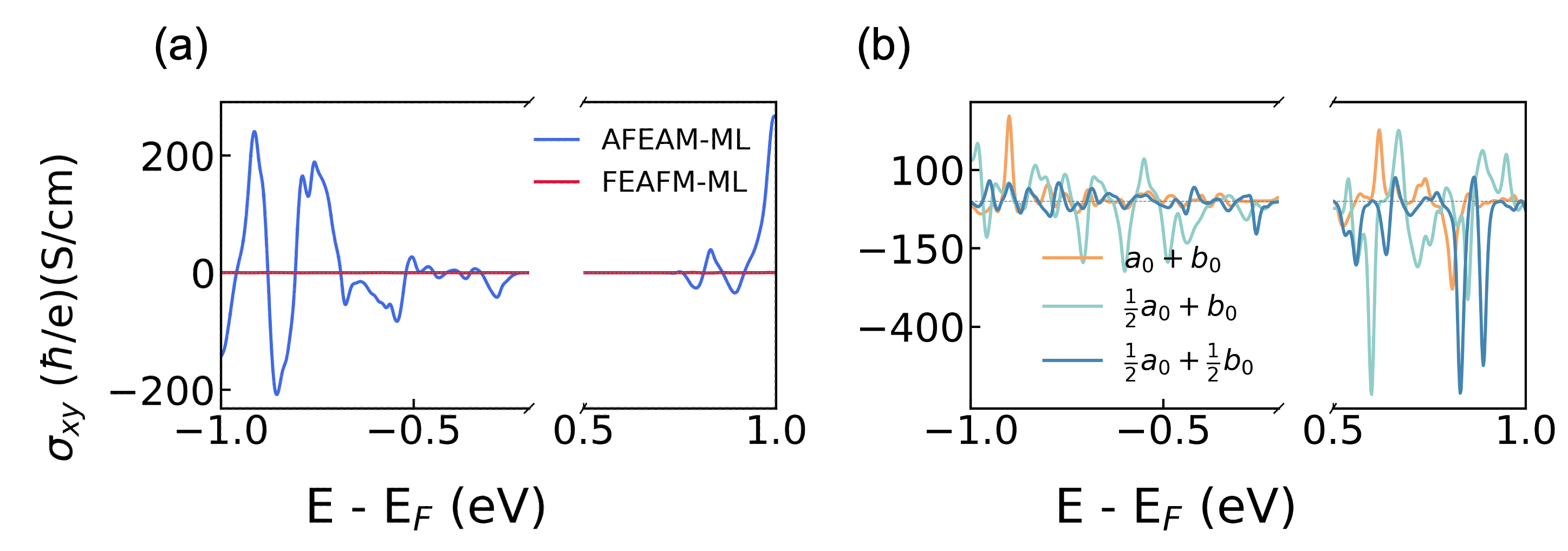}
  \caption{AHC $\sigma_{xy}$ as a function of Fermi energy for (a) AFEAM-ML (blue line) and FEAFM-ML (red line), respectively. (b) AHC for AA-stacked type-II FEAFM-BL FeCuP$_2$S$_6$. The orange, green, and blue lines show the different slding vectors.}
  \label{fig:fig5}  
\end{figure}

To demonstrate how the sliding-driven spin splitting can be experimentally detected, we calculate the anomalous Hall conductivity (AHC, $\sigma_{xy}$) for AFEAM-ML, FEAFM-ML and type-I FEAFM-BL FeCuP$_2$S$_6$. As expected, the FEAFM-ML structure only possesses pure $\tau$ symmetry, resulting in spin-degenerate bands and zero AHC (Figure~\ref{fig:fig5}(a)). The AFEAFM-ML structure, on the other hand, can generate nonzero AHC due to the non-vanishing Berry curvature at the energy with spin splitting. Considering N\'eel vector along [010], the nonzero tensor components of AHE belongs to the MSG $P2_1^{'}$ taking the following form
\[
\begin{pmatrix}
0 & \sigma_{xy} & 0\\
-\sigma_{xy} & 0 & \sigma_{yz} \\
0 & -\sigma_{yz} & 0
\end{pmatrix}
\]
with nonzero values for $\sigma_{xy}$ and $\sigma_{yz}$.
The $\sigma_{xy}$ are all zero at the Fermi level as a result of the band gap and zero net magnetization. However, when going to lower energy, while FEAFM-ML remains zero $\sigma_{xy}$, peaks of $\sigma_{xy}$ are observed for AFEAM-ML. The Berry curvature $\Omega_z$ at the energy (0.3 eV below E$_F$) of the first negative peak (Figure~S11) shows that at some $k$ points near the $\Gamma$ points it does not cancel out, indicating the breaking of time reversal symmetry and thus the nonzero AHE. Figure~\ref{fig:fig5} (b) shows the AHE for three different sliding positions $\Delta S$ of type-II AFEAFM-BL FeCuP$_2$S$_6$. It shows that starting from $\Delta S = a_0 + b_0$, then to $\Delta S = \frac{1}{2}a_0 + b_0$, and further to $\Delta S = \frac{1}{2}a_0 + \frac{1}{2}b_0$, reversal of $\sigma_{xy}$ are observed at specific energies at the three positions with the unchanged N\'eel vector but only different electric polarization due to sliding. 

\section{Conclusion}

In summary, we have discovered ferroelectricity-driven AM in 2D van der Waals multiferroics. In the AFEAFM system, when the magnetic sublattices are connected not merely through pure translation or spatial inversion symmetry, a \textit{k}-dependent spin splitting naturally emerges. We verified this rule in both monolayer (ML) and bilayer (BL) FeCuP$_2$S$_6$ structures. The system exhibits a screw symmetry operation $\{ C_{2y} | 0\frac{1}{2}0 \}$, under which magnetic atoms with opposite spins located along the screw axis give rise to the altermagnetic behavior.
Moreover, the AM state can be preserved during bilayer sliding along the direction perpendicular to the screw/rotation axis ($\vec{b}/b_0$ = 1/2 and 1). The direction and magnitude of the spin-splitting can be tuned by the sliding vector, indicating that the interlayer interaction provides an effective handle to control altermagnetism. Finally, we demonstrate the feasibility of experimentally detecting AM via the anomalous Hall effect. The symmetry-guided design principle proposed here can be extended to other van der Waals magnets and multiferroic systems, offering new opportunities for engineering tunable spin-split states in next-generation spintronic devices.

\footnotesize
\section*{Acknowledgements}
The authors gratefully acknowledge Dr. Ruiwen Xie for insightful discussions. The authors thank the computing time provided to them on the high-performance computer Lichtenberg at the NHR Centers NHR4CES at TU Darmstadt. This work is funded by the Deutsche Forschungsgemeinschaft (DFG, German Research Foundation) - CRC 1487, "Iron, upgraded!" - with project number 443703006. H. Wang and H. Zhang also acknowledge support from the Deutsche Forschungsgemeinschaft (DFG, German Research Foundation) under Project-ID 463184206 – SFB 1548. W. Ren acknowledges support from the National Natural Science Foundation of China (Grant Nos. 52130204, 12311530675), Shanghai Engineering Research Center for Integrated Circuits and Advanced Display Materials, High-Performance Computing Center, Shanghai Technical Service Center of Science and Engineering Computing, Shanghai University. 
\normalsize
\bibliography{references}

\newpage
\section*{Supplement Material}
\setcounter{figure}{0}
\renewcommand{\thefigure}{S\arabic{figure}}
\setcounter{section}{0}
\renewcommand{\thesection}{S\arabic{section}}
\section{Calculation methods}

DFT calculations have been carried out using the projector augmented wave method~\cite{kresse1999ultrasoft}, as implemented in the VASP
package~\cite{kresse1996efficient}. A plane-wave cut-off energy of 500~eV and a 8x5x1 k-point mesh in the irreducible Brillouin zone were used in the calculations.  The Perdew-Burke-Ernzerhof (PBE)~\cite{perdew1996generalized} functional within the generalized-gradient approximation, along with a Hubbard-U correction, was employed to accurately
describe electronic interactions. Following Refs.~\cite{wang2025magnetic,yang2024understanding}, an effective U value of 3.0 eV was applied to Fe-\textit{d} orbitals to account for the on-site Coulomb repulsion. Grimme's D3 correction~\cite{grimme2011effect} was used to include the van-der-Waals interactions. Nudged elastic band (NEB) was applied in the computation of the transition pathway between ferroelectric, antiferroelectric, and paraelectric configurations. 

\subsection{Anomalous Hall conductivity (AHC)}
The intrinsic component of anomalous Hall conductivity \(\sigma_{xy}\), which arises solely from the electronic band structure, can be directly evaluated using the Kubo formula~\cite{greenwood1958boltzmann}:

\begin{equation}
\sigma_{xy} = -\frac{e^2}{\hbar} \int_{\text{BZ}} \frac{d\mathbf{k}}{(2\pi)^3} 
\sum_n f(\varepsilon_{n\mathbf{k}}) \Omega_{n,xy}(\mathbf{k}).
\label{eq:Kubo}
\end{equation}
Here, \(\hbar\), \(e\), and \(\varepsilon_{n\mathbf{k}}\) represent the reduced Planck constant, 
positive elementary charge, and eigenenergy. The Fermi distribution function is denoted by 
\(f(\varepsilon) = (e^{(\varepsilon - \mu)/k_B T} + 1)^{-1}\), 
where \(\mu\) stands for the chemical potential. 
The Berry curvature \(\Omega_{n,xy}(\mathbf{k})\) for band \(n\), which can be expressed as,
\begin{equation}
\Omega_{n,xy}(\mathbf{k}) = -2\hbar^2 \mathrm{Im} 
\sum_{m \ne n} 
\frac{
\langle n\mathbf{k} | \hat{v}_x | m\mathbf{k} \rangle
\langle m\mathbf{k} | \hat{v}_y | n\mathbf{k} \rangle
}{
(\varepsilon_{n\mathbf{k}} - \varepsilon_{m\mathbf{k}})^2
}.
\label{eq:berry}
\end{equation}
with \(\hat{v}_x\) (\(\hat{v}_y\)) being the \(k_x\) (\(k_y\)) component of the velocity operator, 
and \(|n\mathbf{k}\rangle\) representing the eigenstate.

\subsection{(Spin) shift current}

The shift current $J_a$ is a second order response and thus can be expressed in terms of two electric field components and material-dependent response function:
\begin{equation}
J_a = \sigma_{abc} E_b E_c
\end{equation}
The shift current spectra are expressed by:
\begin{equation}
\sigma^{abc}(\omega) = \frac{i \pi e^3}{2 \hbar^2} 
\int \frac{d k}{8 \pi^3} 
\sum_{n,m} \left( r_{mn}^b r_{nm;a}^c + r_{mn}^c r_{nm;a}^b \right) 
\delta(\omega_{mn} - \omega)
\end{equation}
where indices $a$, $b$, and $c$ represent Cartesian directions, 
$r_{mn}^b$ denotes the velocity matrix elements, 
and $r_{nm;a}^c$ denotes the generalized derivatives, which are defined as:
\begin{equation}
r_{nm;a}^c = \frac{\partial r_{nm}^c}{\partial k_a} 
- i \left( A_{mm}^a - A_{nn}^a \right) r_{nm}^c
\end{equation}
with the Berry connections $A_{mm}^a$. Maximally localized Wannier functions to fit DFT electronic structures are obtained using the Wannier90 code ~\cite{mostofi2008wannier90} with projection of Fe-$d$, Cu-$d$, P-$p$, and S-$p$ orbitals. To calculate the AHC using WANNIER90 code, a k mesh of $120 \times 120 \times 1$ was taken in the integration for AHC to address the rapid variation of Berry curvature. The WannierBerri code~\cite{tsirkin2021high} was used to evaluate the shift current tensor with $200 \times 200 \times 1$ k-mesh.

\section{Crystal structure}

\begin{figure}[H]
  \centering
  \includegraphics[width=0.48\textwidth]{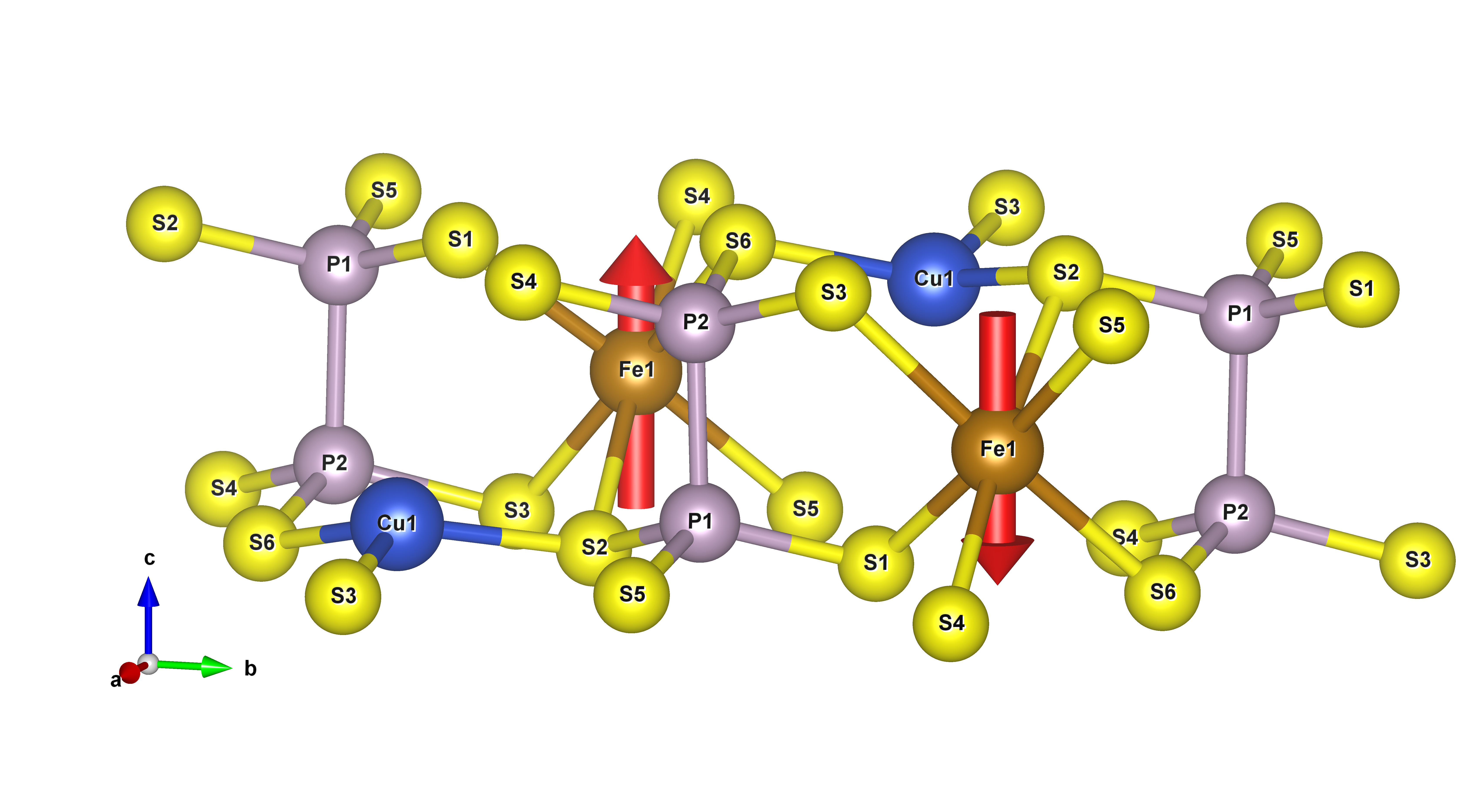}
  \caption{Crystal structure of AFEAFM-ML FeCuP$_2$S$_6$. Lattice constants and the corresponding atomic positions are listed in Table~S1.}
  \label{fig:strcture}  
\end{figure}

\begin{table*}[t]
  \caption{Lattice constants, Wyckoff positions, bond lengths and magnetic moments of magnetic atoms for ML and BL FeCuP$_2$S$_6$.}
  \centering
  \begin{tabular}{lllll}
    \toprule
    Material  &  Lattice constants  &  Wyckoff positions  & Bond lengths [\AA] & Magnetic moments [$\mu_B$] \\
    \midrule
    FeCuP$_2$S$_6$ (4) & \makecell{a = 5.97\AA \\ b = 10.34 \AA \\ c = 30.37\AA \\ $\alpha$ = $\beta$ = $\gamma$ = 90.0\textdegree}  & \makecell{Fe1 2a (0.243, 0.333, 0.502) \\ Cu1 2a (0.750, 0.165, 0.459) \\ P1 2a (0.242, 0.003, 0.534) \\ P2 2a (0.242, 0.003, 0.534) \\ S1 2a (0.399, 0.168, 0.551) \\ S2 2a (0.560, 0.347, 0.447) \\ S3 2a (0.114, 0.175, 0.444) \\ S4 2a (-0.065, 0.340, 0.555) \\ S5 2a (-0.083, -0.010, 0.551) \\ S6 2a (0.574, -0.019, 0.443)}  & \makecell{Fe-S (2.45, 2.46, 2.51) \\ Cu-S (2.22) \\ P-P (2.20) \\ P-S (2.01, 2.07) } & \makecell{Fe 3.84, -3.84 \\ Cu -0.002, 0.002 \\ S1 0.076, -0.076 \\ S2 0.061 -0.061 \\ S3 0.067 -0.067 \\ S4 0.081, -0.081 \\ S5 -0.1, 0.1 \\ S6 -0.082, 0.082} \\
    BL FeCuP$_2$S$_6$ (4) & \makecell{a = 5.97\AA \\ b = 10.34 \AA \\ c = 34.12\AA \\ $\alpha$ = $\beta$ = $\gamma$ = 90.0\textdegree}  & & & \makecell{type-I \\ Fe 3.833, -3.833 \\ type-II \\ Fe 3.85, -3.85 \\ Cu 0.083, -0.083 \\ type-III \\Fe 3.833, -3.833 }  \\
    \bottomrule
  \end{tabular}
  \label{tab:table}
\end{table*}

\newpage

\subsection{Electronic properties}

\begin{figure}[H]
  \centering
  \includegraphics[width=0.48\textwidth]{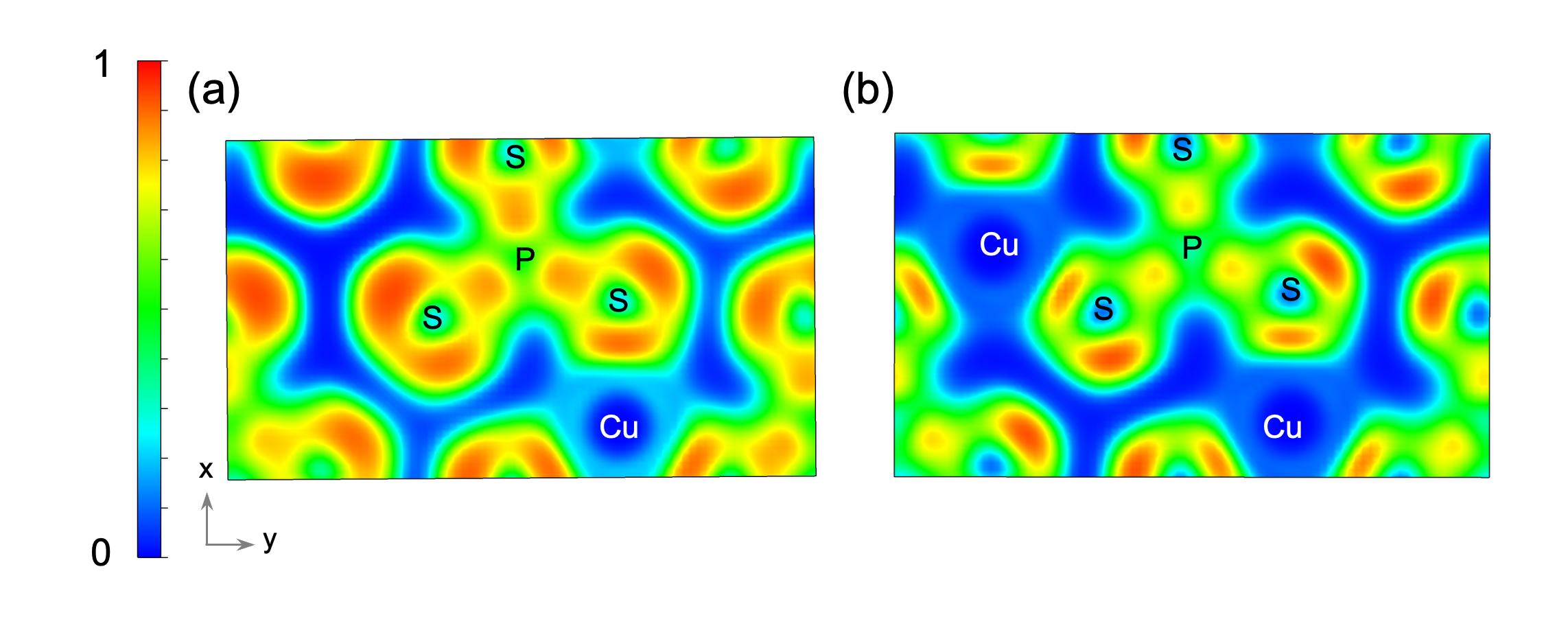}
  \caption{Electronic localization function for (a) AFEAFM and (b) FEAFM ML FeCuP$_2$S$_6$ at z=14.3 \AA\ plane. ELF = 1 (red) and 0 (blue)  indicate accumulated and vanishing electron density, respectively.}
  \label{fig:elf}  
\end{figure}


\begin{figure}[H]
  \centering
  \includegraphics[width=0.48\textwidth]{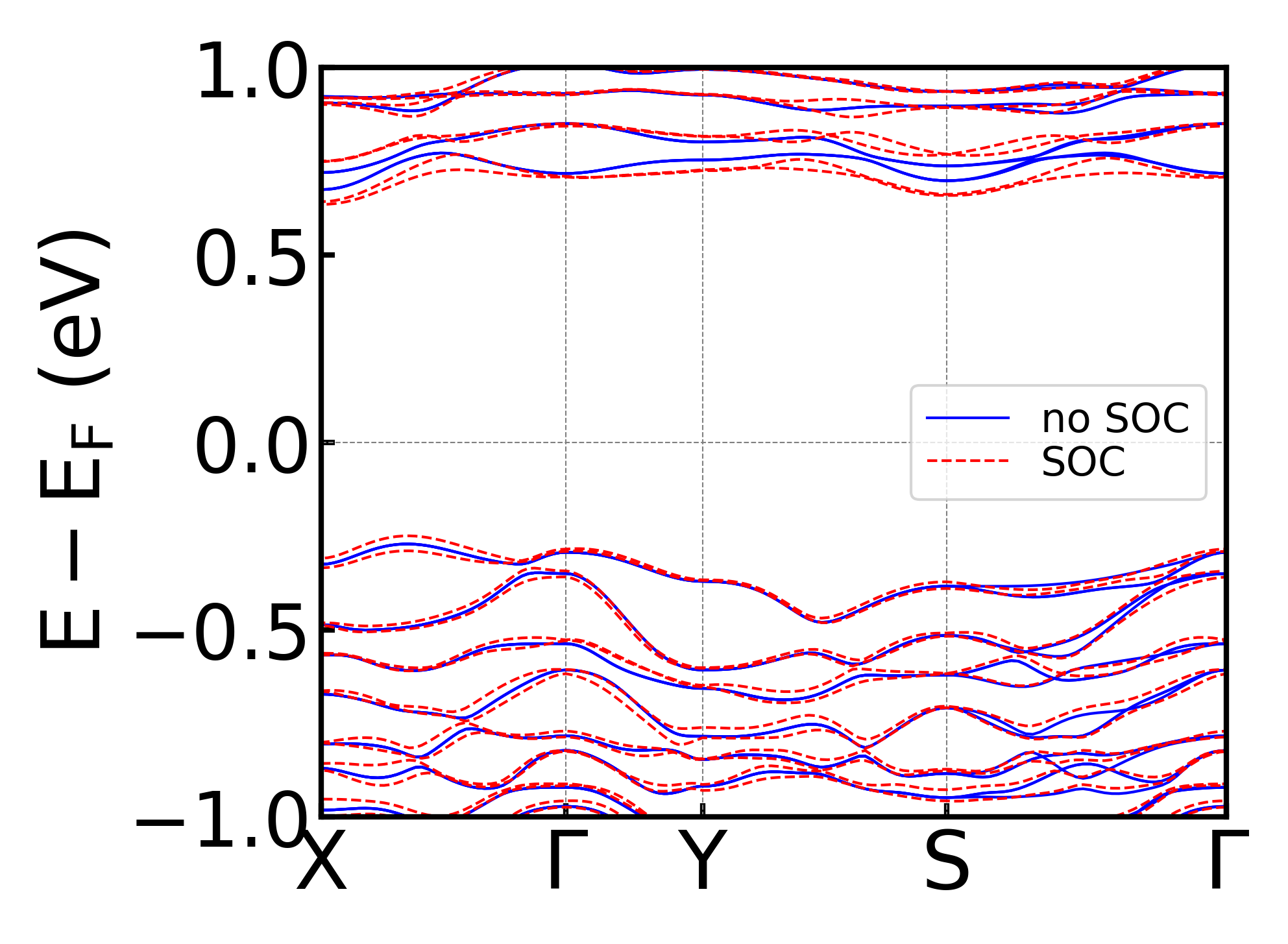}
  \caption{Band structure for ML-FeCuP$_2$S$_6$ with (blue line) and without (red dashed line) SOC.}
  \label{fig:bands_soc}  
\end{figure}


\begin{figure}[H]
  \centering
  \includegraphics[width=0.48\textwidth]{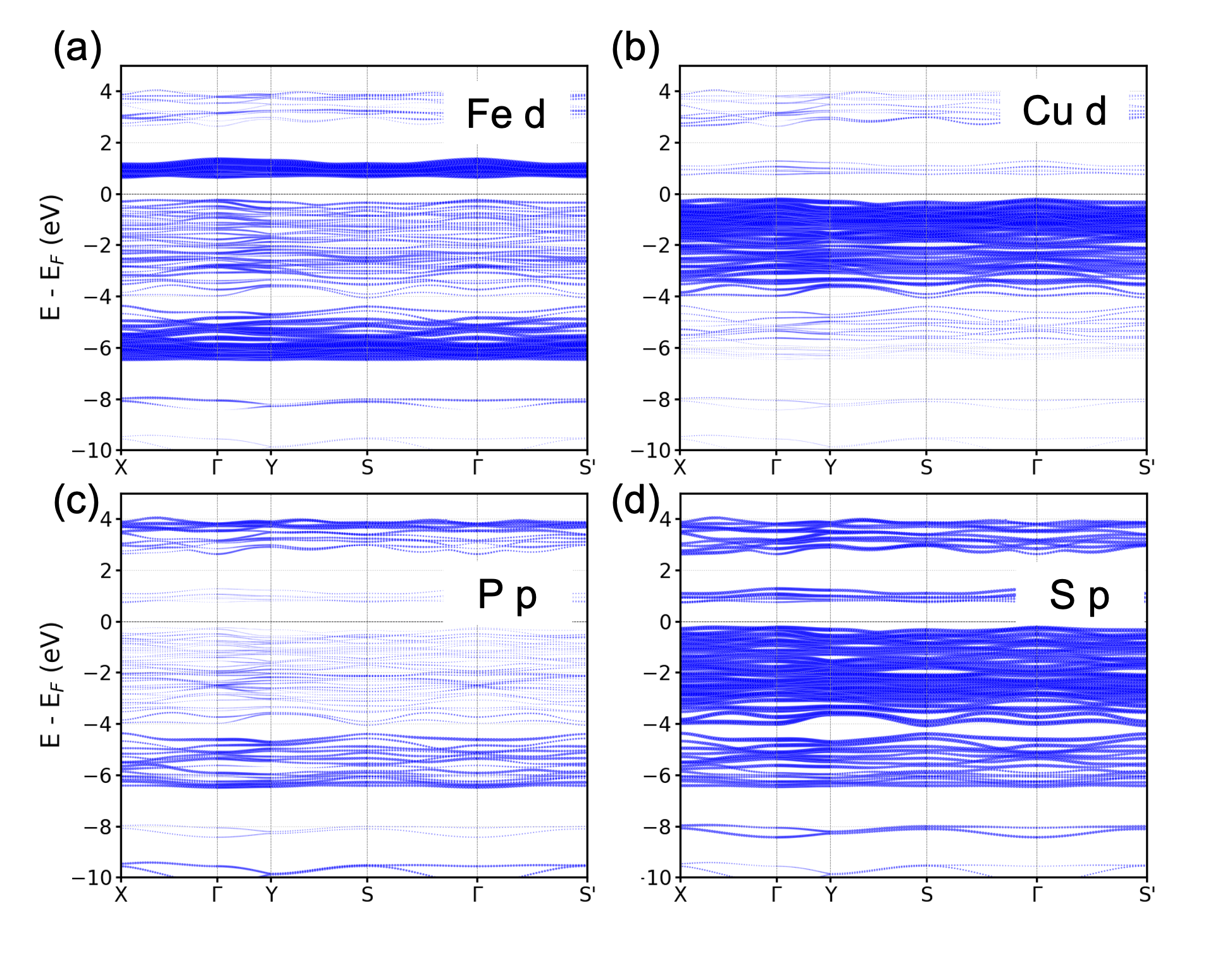}
  \caption{Fat bands of ML FeCuP$_2$S$_6$: (a) d orbitals of Fe, (b) d orbitals of Cu, (c) p orbitals of P, and (d) p orbitals of S.}
  \label{fig:fat_band}  
\end{figure}


\begin{figure}[H]
  \centering
  \includegraphics[width=0.48\textwidth]{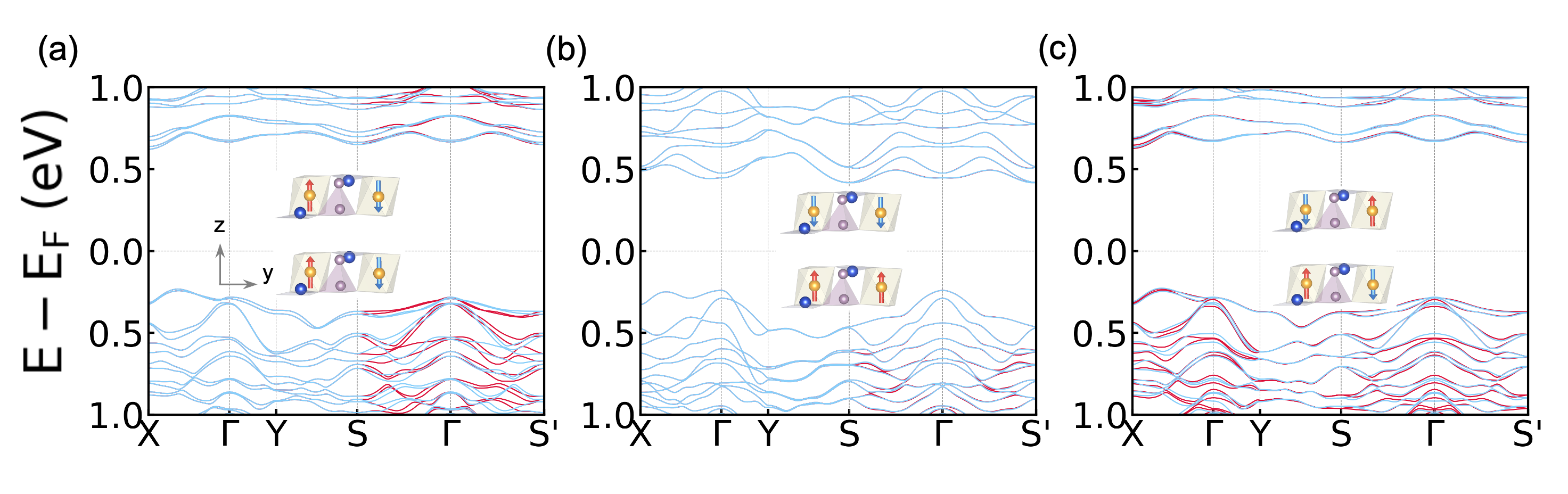}
  \caption{Band structure for (a) type-I, (b) type-II, and (c) type-III stacked AFEAM-BL FeCuP$_2$S$_6$. Insets show the crystal structures and magnetic configurations.}
  \label{fig:bands}  
\end{figure}

\begin{figure}[H]
  \centering
  \includegraphics[width=0.48\textwidth]{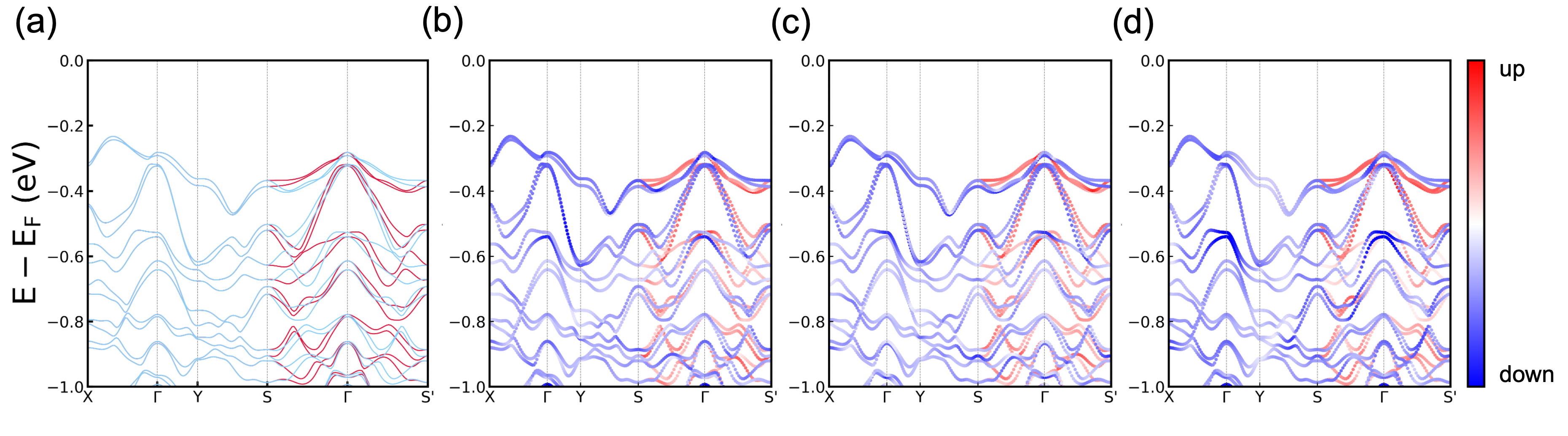}
  \caption{Type-I BL FeCuP$_2$S$_6$ exhibits AFM in each layer. the projected band structure of upper-layer Fe atoms, lower-layer Fe atoms and one pair of screw axis equivalent Fe atoms, respectively, shown in (b), (c), and (d).  Blue and red lines indicate spin-up and spin-down channels, respectively.}
  \label{fig:udud}  
\end{figure}

To figure out whether the spin splitting is controlled only by the screw axis symmetry or it involves also the interlayer interactions or other effect, we plot the projected band structure of upper-layer Fe atoms, lower-layer Fe atoms and one pair of screw axis equivalent Fe atoms, respectively, shown in (b), (c), and (d). The total band structure is shown in (a) as reference. Spin splitting is observed in all the three cases, indicating that it is influenced by both the screw symmetry and the interlayer interactions.

\begin{figure}[H]
  \centering
  \includegraphics[width=0.48\textwidth]{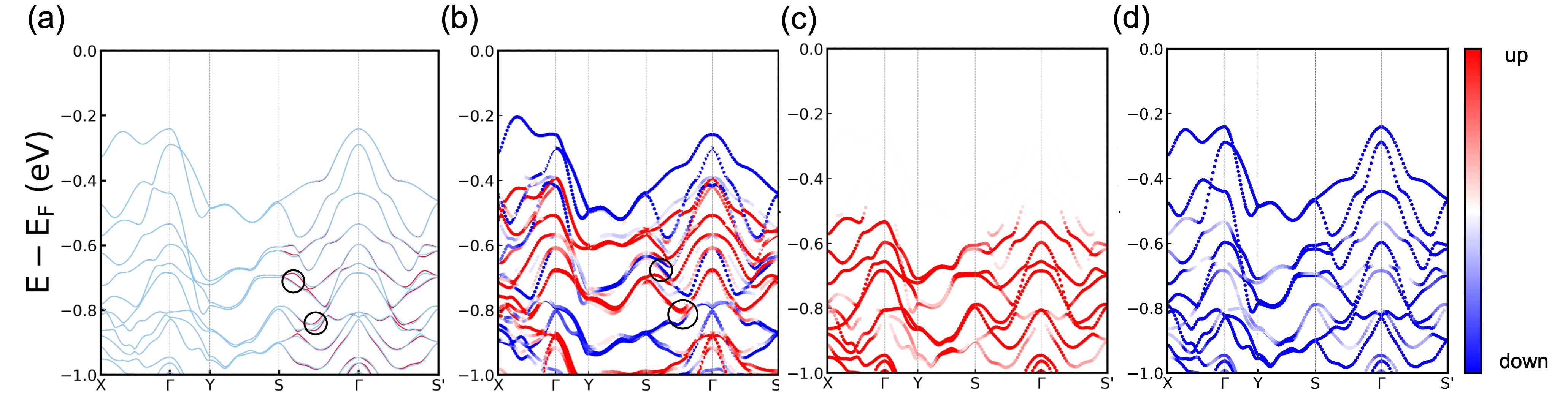}
  \caption{In type-II BL FeCuP$_2$S$_6$ it is found that the Cu also contributes to the the magnetism with magnetic moment 0.085 and -0.085 $\mu_B$ for the Cu atoms in the upper and the lower layer, respectively. In order to explore in detail the role of Cu in the generation of AM, we calculate the band structure with magnetic moment of Cu fixed at zero. The result is shown in (b), as compared to the initial spin-polarized band structure shown in (a). Altermagnetism-induced spin splitting still exists when Cu atoms show zero magnetic moment (black circles highlight two of the split bands), indicating that like type-I BL FeCuP$_2$S$_6$, Fe atoms still contribute to the spin splitting in type II stacking. Therefore, the general symmetry rule of the generation of AM in BL FeCuP$_2$S$_6$ discussed in the main manuscript still holds. (c) and (d) show the projected band structure of the upper layer Fe and the lower layer Fe of type-II FeCuP$_2$S$_6$, respectively.}
  \label{fig:uudd}  
\end{figure}


\begin{figure}[H]
  \centering
  \includegraphics[width=0.48\textwidth]{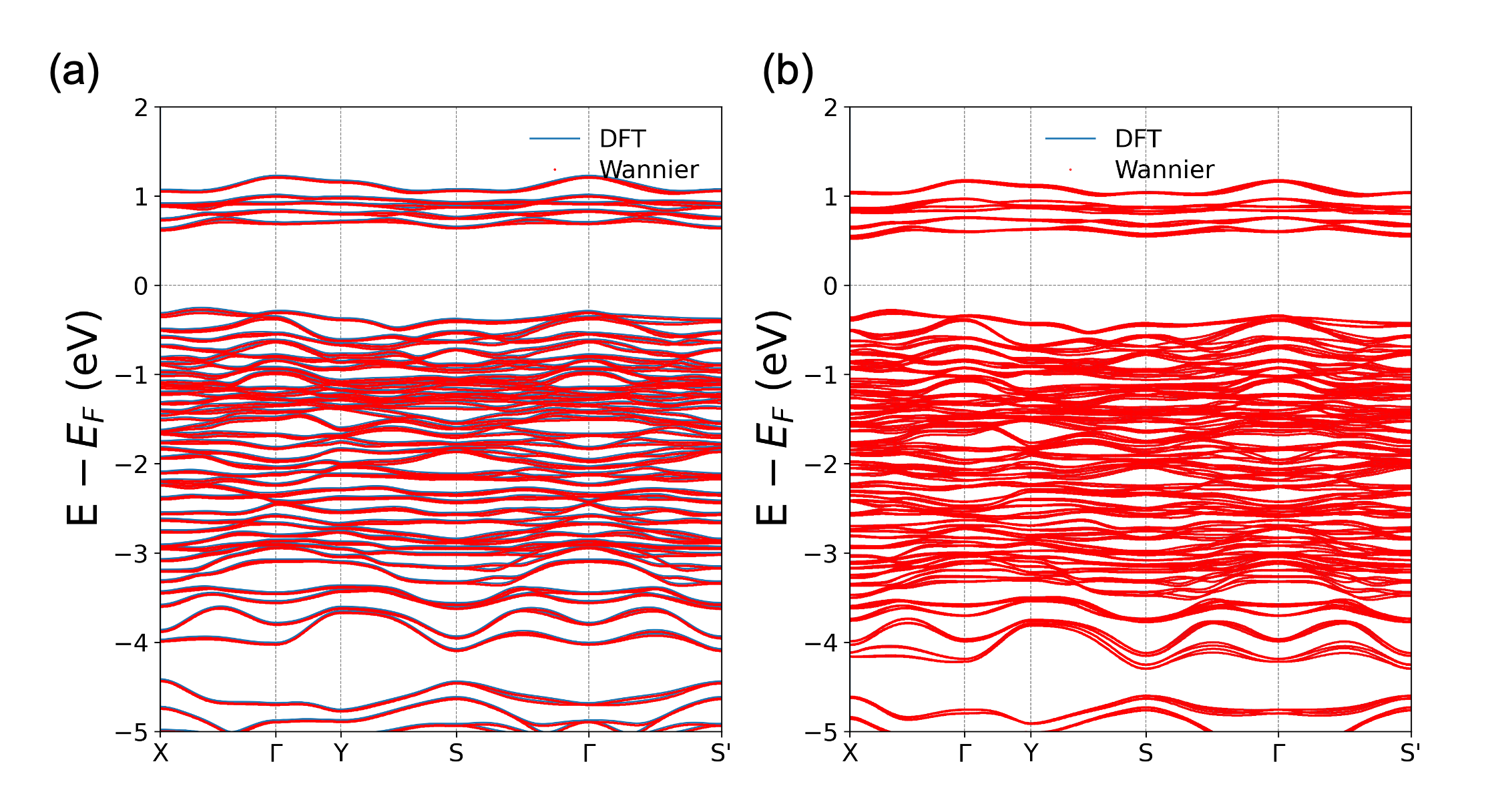}
  \caption{DFT (blue lines) and Wannier (red dots) band structure for (a) AFEAM-ML and (b) AFEAM-BL FeCuP$_2$S$_6$.}
  \label{fig:dft_wann}  
\end{figure}


\section{Berry curvature related properties}

\begin{table}[h]
\caption{Matrix elements of shift current tensor for AFEAFM-ML FeCuP$_2$S$_6$.}
\centering
\[
\begin{pmatrix}
0 & 0 & 0 & \sigma_{xyz} & 0 & \sigma_{xxy} \\
\sigma_{yxx} & \sigma_{yyy} & \sigma_{yzz} & 0 & \sigma_{yxz} & 0 \\
0 & 0 & 0 & \sigma_{zyz} & 0 & \sigma_{zxy}
\end{pmatrix}
\]
\end{table}

\begin{figure}[H]
  \centering
  \includegraphics[width=0.48\textwidth]{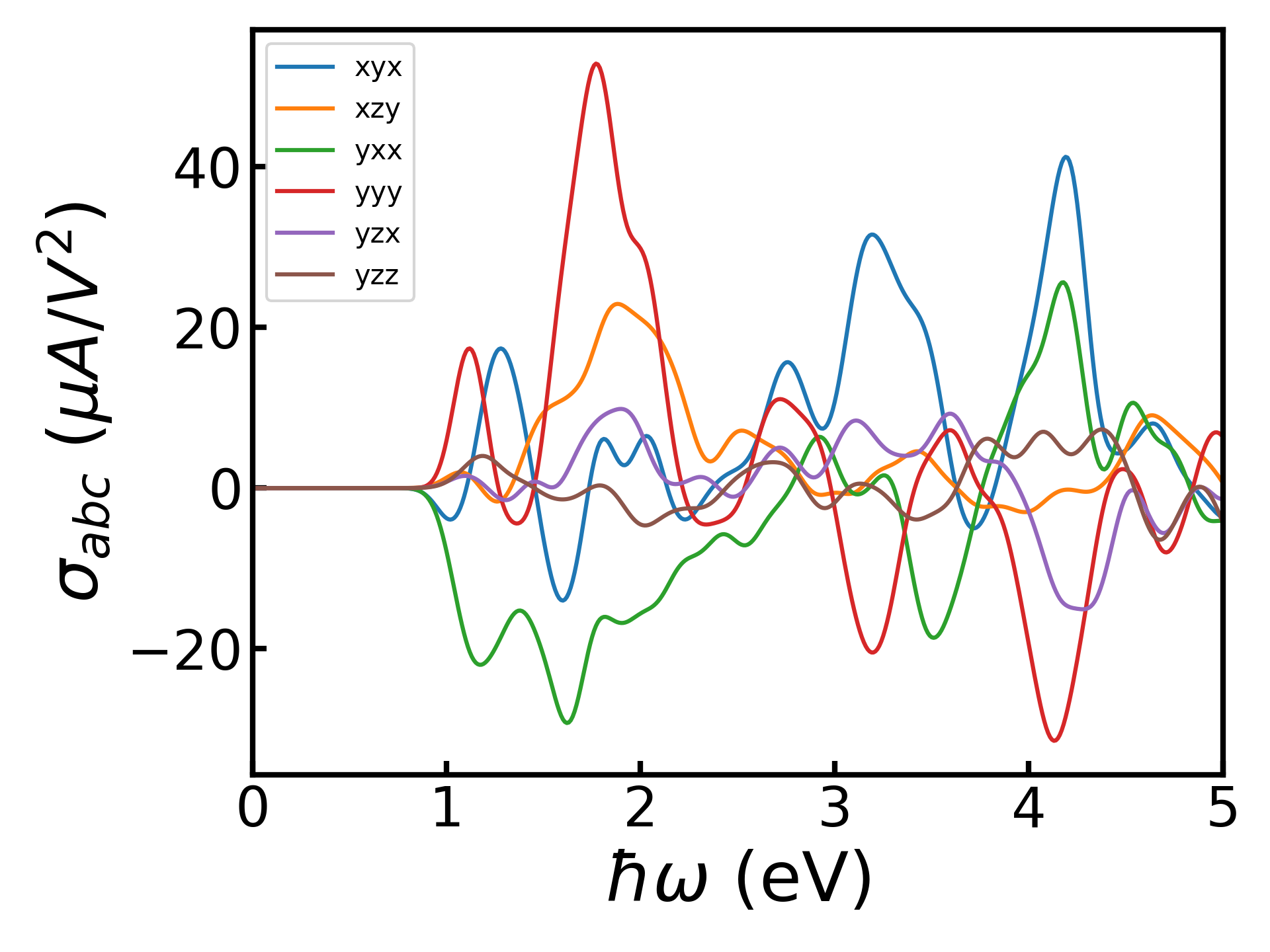}
  \caption{Shift current calculated for FEAFM-ML FeCuP$_2$S$_6$. The x-axis stands for the photon energy. Different colors represent different components of the $\sigma_{abc}$ tensor.}
  \label{fig:shift_current}  
\end{figure}

\begin{figure*}[t]
  \centering
  \includegraphics[width=0.67\textwidth]{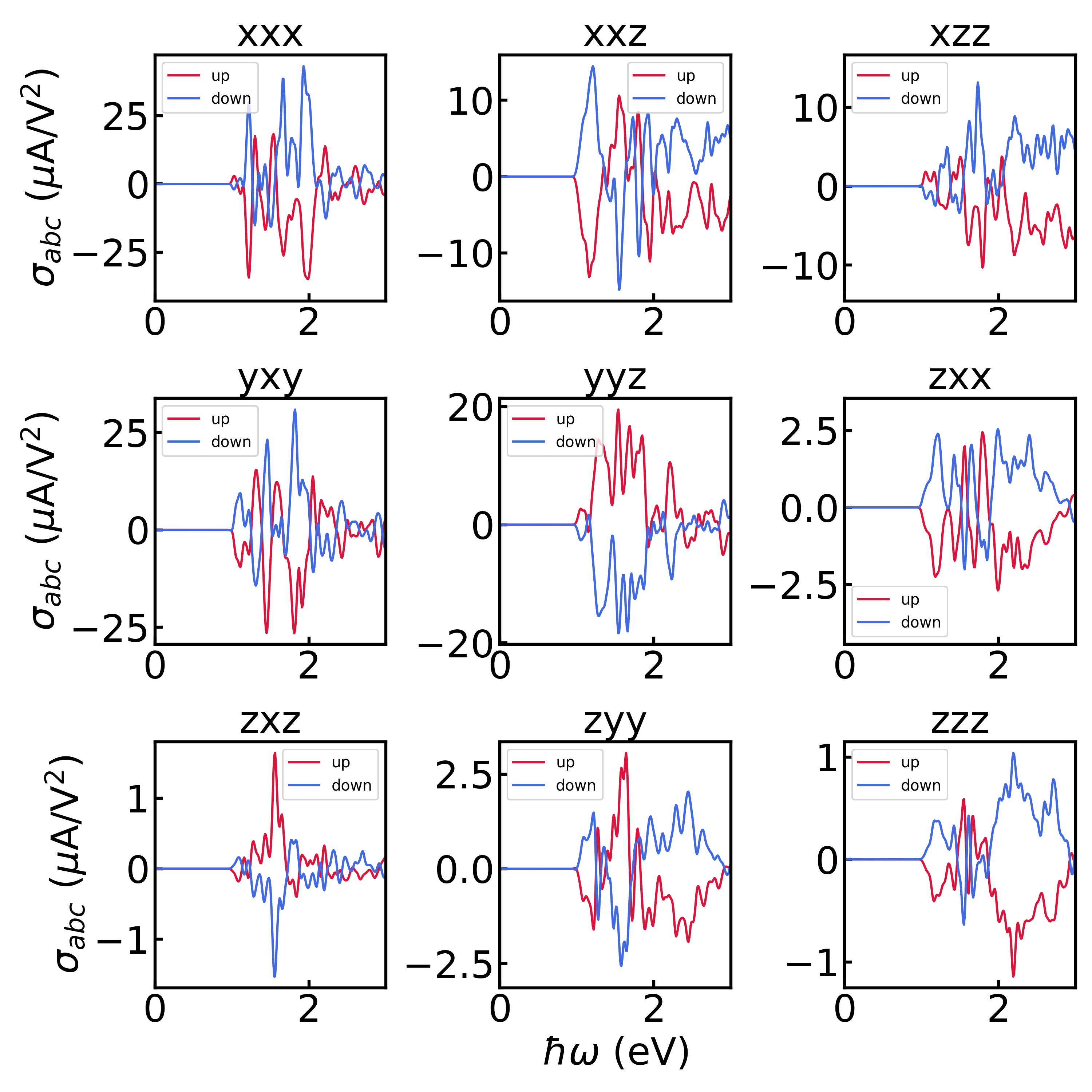}
  \caption{Non-vanishing spin current components of FEAFM-ML FeCuP$_2$S$_6$ with separated spin up and down channels.}
  \label{fig:spin_current}  
\end{figure*}


\begin{figure}[H]
  \centering
  \includegraphics[width=0.48\textwidth]{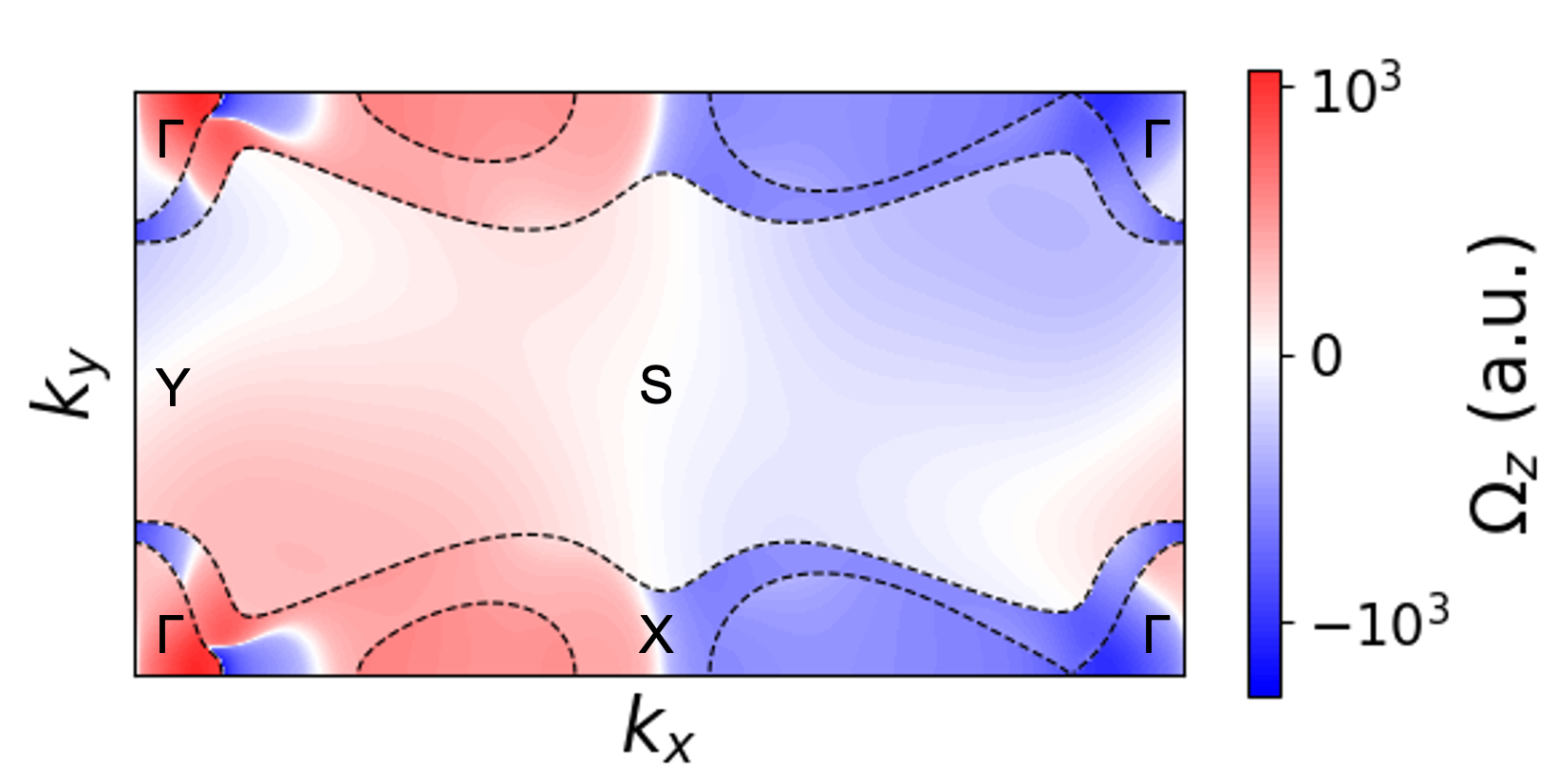}
  \caption{Berry curvature of AFEAM-ML FeCuP$_2$S$_6$ at the energy of 0.3 eV below E$_F$ where the anomalous Hall conductivity $\sigma_{xy}=-50$ ($\hbar/e$)(S/cm) (corresponding to the first negative peak below E$_F$ observed in Figure~5 (a). Red and blue represent positive and negative Berry curvature.}
  \label{fig:berry}  
\end{figure}

\end{document}